\documentclass[twocolumn,showpacs,preprintnumbers, prc,superscriptaddress]{revtex4}
\usepackage{epsfig}
\usepackage{graphicx}
\usepackage{epstopdf}
\usepackage{amsmath,amssymb,amsfonts}
\usepackage{array}
\usepackage{url}
\usepackage{hyperref}
\usepackage{lineno}
\usepackage{xspace}
\usepackage[usenames,dvipsnames]{color}
\newcommand{\sqsn}{\mbox{$\sqrt{s_{_{NN}}}$}\xspace}
\newcommand{\bef}{\begin{figure}}
\newcommand{\eef}{\end{figure}}
\newcommand{\bc}{\begin{center}}
\newcommand{\ec}{\end{center}}

\begin{document}
\title{Resonance decay effect on conserved number fluctuations in a hadron 
resonance gas model}

\author{D.~K.~Mishra}
\email {dkmishra@rcf.rhic.bnl.gov}
\affiliation{Nuclear Physics Division, Bhabha Atomic Research Center, Mumbai 
400085, India}
\author{P.~Garg}
\affiliation{Department of Physics and Astronomy, Stony Brook University, SUNY, 
Stony Brook, New York 11794-3800, USA}
\author{P.~K.~Netrakanti}
\affiliation{Nuclear Physics Division, Bhabha Atomic Research Center, 
Mumbai 400085, India}
\author{A.~K.~Mohanty\footnote{Presently at Saha Institute of Nuclear Physics, 
1/AF, Bidhan nagar, Kolkata - 700064, India}}
\affiliation{Nuclear Physics Division, Bhabha Atomic Research Center, Mumbai 
400085, India}

\begin{abstract}
We study the effect of charged secondaries coming from resonance decay on the 
net-baryon, net-charge and net-strangeness fluctuations in high energy heavy-ion 
collisions within the hadron resonance gas (HRG) model. We emphasize the 
importance of including weak decays along with other resonance decays in the 
HRG, while comparing with the experimental observables. The effect of kinematic 
cuts on resonances and primordial particles on the conserved number fluctuations 
are also studied. The HRG model calculations with the inclusion of resonance 
decays and kinematical cuts are compared with the recent experimental data from 
STAR and PHENIX experiments. We find a good agreement between our model 
calculations and the experimental measurements for both net-proton and 
net-charge distributions.

\end{abstract}

\maketitle

\section{Introduction}
\label{intro} 
Beam Energy Scan (BES) program at Brookhaven National Laboratory's Relativistic 
Heavy-Ion Collider (RHIC) has drawn much attention to explore the quantum 
chromodynamics (QCD) phase diagram in terms of temperature ($T$) 
and baryon chemical potential 
($\mu_{B}$)~\cite{Stephanov:1998dy,Alford:1997zt,Stephanov:1996ki,Aoki:2006we}. 
Several theoretical models suggest the existence of a critical point in the 
$T-\mu_B$ plane where the first order phase transition line originating from 
high $\mu_{B}$ ends~\cite{Stephanov:2004wx,Fodor:2004nz,Stephanov:1999zu}. 
The location of the critical point can be explored by systematically varying 
$T$ and $\mu_B$. Experimentally, one can vary the $T$ and $\mu_B$ by varying 
the center of mass energy of the colliding ions. It has been suggested that the 
excitation function of conserved numbers like net-baryon, net-charge and 
net-strangeness fluctuations should show a non-monotonic behavior, as a possible 
signature of QCD critical end point 
(CEP)~\cite{Koch:2005vg,Asakawa:2000wh,Asakawa:2009aj}.

In the thermodynamic limit, the correlation length ($\xi$) diverges at 
CEP~\cite{Stephanov:1998dy}. The moments of the net-baryon, net-charge, and 
net-strangeness distributions are related to the $\xi$ of the system and hence 
these moments can be used to look for signals of a phase transition and 
critical point~\cite{Ejiri:2005wq,Bazavov:2012vg}. Also, the comparison of 
experimentally measured cumulants with the lattice calculations enables us to 
extract the freeze-out parameters i.e. freeze-out temperature ($T_f$) and 
$\mu_B$ of the system produced in heavy-ion 
collisions~\cite{Borsanyi:2014ewa,Alba:2014eba,Adare:2015aqk}. In recent years, 
lots of efforts have been put on both theoretical and experimental fronts to 
study the fluctuation of conserved quantities. Current experiments at RHIC 
(PHENIX and STAR), have reported their measurements of higher moments for 
net-charge~\cite{Adare:2015aqk,Adamczyk:2014fia} and for 
net-proton~\cite{Adamczyk:2013dal} multiplicity distributions at 
different center-of-mass energies (\sqsn) using the data from BES. 
Experimentally, the net-baryon number fluctuations are not directly measured, as 
all the neutral baryons are not detected by most of the experiments. Hence, 
net-baryon fluctuations are accessible via measuring the net-proton 
distributions~\cite{Aggarwal:2010wy}. The net-charge fluctuations are accessible 
by measuring the stable charged particles such as pions, kaons and protons 
along with their anti particles~\cite{Adare:2015aqk,Adamczyk:2014fia}. 
Similarly, the measurement of net-kaon acts as a proxy for 
net-strangeness fluctuations, as higher mass strange particles are not 
directly measured. There are several sources of non-equilibrium fluctuations, 
which can diminish the fluctuations measured by the 
experiments~\cite{Berdnikov:1999ph}. It is important to identify the 
non-critical baseline to understand the critical properties of different 
conserved number fluctuations. 

Experimentally measured moments of the net distributions are related to the 
cumulants as: mean ($M$) = $C_1$; variance $\sigma^2 = C_2 = \langle (\delta 
N)^2\rangle$; skewness $S = C_3/C_2^{3/2} = \langle (\delta 
N)^3\rangle/\sigma^3$ and kurtosis $\kappa = C_4/C_2^2 = \langle (\delta 
N)^4\rangle/\sigma^4 - 3$, where $N$ is the multiplicity of the distribution and 
$\delta N$ = $N-M$. Hence, the ratios of the cumulants are related to the 
moments as: $\sigma^2/M = C_2/C_1$, $S\sigma = C_3/C_2$, $\kappa\sigma^2 = 
C_4/C_2$ and $S\sigma^3/M = C_3/C_1$. Further, the ratios of moments/cumulants 
can be related to the susceptibilities of $n$th order ($\chi^n$) obtained from 
the lattice QCD or the HRG model calculations as $\sigma^2/M \sim 
\chi^{(2)}/\chi^{(1)}$, $S\sigma \sim \chi^{(3)}/\chi^{(2)}$, $\kappa\sigma^2 
\sim \chi^{(4)}/\chi^{(2)}$, and $S\sigma^3/M \sim \chi^{(3)}/\chi^{(1)}$. One 
advantage of measuring the ratios is that the volume dependence on the 
experimental measured individual cumulants cancel out in the ratios. 
Experimentally, one measures only the final abundance of hadrons which 
includes both primordial particle production as well as contributions from the 
resonance decays. Production of resonances play important role for studying 
various properties of interaction dynamics in the heavy-ion collisions. 
Resonances having short life time which subsequently decays into stable hadrons 
and can affect the final hadron yields and their number fluctuations.  

The HRG model has been successful in explaining the particles produced in 
heavy ion collisions from AGS to LHC 
energies~\cite{BraunMunzinger:2003zd,Cleymans:2005xv,Andronic:2011yq}. The 
susceptibilities and their ratios in hadronic phase calculated in the HRG model 
reasonably agrees with the lattice QCD results at lower 
$\mu_B$~\cite{Bazavov:2012vg}. Several studies have been performed with the HRG 
model for the fluctuation of conserved quantities, which are considered as 
baseline for such 
measurements~\cite{Begun:2006jf,Karsch:2010ck,Garg:2013ata,Fu:2013gga,
Rau:2013xya}. 
Also, similar baseline studies have been performed using independent production 
model and transport 
model~\cite{Netrakanti:2014mta,Mishra:2015ueh,Tarnowsky:2012vu,Westfall:2014fwa}
. Keeping in mind the existence of CEP in the QCD phase diagram and the efforts 
put in both theoretical and experimental side, it is important to calculate the 
more appropriate baseline for comparison with the experimental data. In the 
present work, we estimate the contribution of resonances to the conserved 
number fluctuation using the HRG model. As discussed in 
Ref.~\cite{Nahrgang:2014fza}, we also calculate the effect of average decay and 
by taking the higher order correlated terms, inclusion of weak decays in the 
model, the effect of different kinematic cuts on the resonance as well as on the 
primordial particles. It is important to consider the weak decays as many of the 
particles which are considered as stable in Ref.~\cite{Nahrgang:2014fza} decay 
before reaching the detector. For example, experimentally $\eta^0 $ or 
$\Lambda^0$ particles are detected by reconstructing from their decayed 
daughters which are measured by the detector. Hence their contributions to the 
fluctuation of stable particles get influenced. 

The paper is organized as follows: In the following section, we discuss the HRG 
model used in this study as well as the implementation of resonance decays. In 
Sec. \ref{sec:results}, the results for the observables 
$\chi^{(2)}/\chi^{(1)}$, $\chi^{(3)}/\chi^{(2)}$, $\chi^{(4)}/\chi^{(2)}$, and 
$\chi^{(3)}/\chi^{(1)}$ for considering different decay cases as well as 
inclusion of weak decays and effect of kinematical cuts are discussed. 
Section~\ref{sec:compare} discuss the comparison of our model calculations to 
the experimental measurements. Finally, in Sec.~\ref{sec:summary} we summarize 
our findings and mention about the implication of this work.

\section{Hadron resonance gas model and resonance decays}
\label{sec:hrg}
The partition function in the HRG model includes all relevant degrees of 
freedom of the confined, strongly interacting matter and contains all the 
interactions that result in resonance formation~\cite{Karsch:2010ck}. The 
heavy-ion experiments cover only a limited phase space, hence one can access 
only a part of the fireball produced in the collision which resemble with the 
Grand Canonical Ensemble~\cite{Koch:2008ia}. Assuming a thermal system produced 
in the heavy-ion collisions, the thermodynamic pressure ($P$) can be written as 
sum of the partial pressures of all the particle species $i$ which may be baryon 
($B$) or meson ($M$).
\begin{equation}
 \frac{P}{T^4} =  \frac{1}{VT^3}\sum_B \mathrm{ln} Z_i(T,V,\mu_i) + 
\frac{1}{VT^3}\sum_M \mathrm{ln} Z_i(T,V,\mu_i)   
\end{equation}
where 
\begin{equation}
 \mathrm{ln} Z_i(T, V, \mu_i) = \pm \frac{Vg_i}{2\pi^2}\int d^3p 
~\mathrm{ln}\big\{1\pm \mathrm{exp}[(\mu_i-E)/T]\big\},
\end{equation}
here $T$ is the temperature, $V$ is the volume of the system, $\mu_{i}$ is the 
chemical potential and $g_{i}$ is the degeneracy factor of the $i$-th 
particle. The total chemical potential of the individual particle is $\mu_i$ = 
$B_i\mu_{B} + Q_i\mu_{Q} + S_i\mu_{S}$, where $B_i$, $Q_i$ and $S_i$ are the 
baryon, electric charge  and strangeness number of the $i$-th particle, with 
corresponding chemical potentials $\mu_{B}$, $\mu_{Q}$ and $\mu_{S}$, 
respectively. The $+ve$ and $-ve$ signs correspond to baryons and mesons 
respectively. In a static fireball, a particle of mass $m$, the volume element 
($d^3p$) can be written as $d^3p = p_Tm_T~\mathrm{cosh}\eta ~dp_T~d\eta 
~d\phi$ and 
energy ($E = m_T\mathrm{cosh}\eta$) of the particle, where $m_T$ is the 
transverse mass $= \sqrt{m^2 + p_T^2}$ with $p_T$, $\eta$ and $\phi$ represents 
the transverse momentum, pseudo-rapidity and azimuthal angle, respectively. The 
experimental acceptances can be applied by considering the corresponding ranges 
in $p_T$, $\eta$ and $\phi$. The fluctuations of the conserved numbers are 
obtained from the derivative of the thermodynamic pressure with respect to the 
corresponding chemical potentials $\mu_B, \mu_Q$ or $\mu_S$. The $n$-th order 
generalized susceptibilities ($\chi$) are written as;
\begin{equation}
 \chi_x^{(n)} = \frac{d^n[P(T,\mu)/T^4]}{d(\mu_x/T)^n}.    
\end{equation}
For mesons $\chi_x$ can be expressed as 
\begin{eqnarray}
 \chi_{x,meson}^{(n)}=\frac{X^n}{VT^3}
\int{d^{3}p}\sum_{k=0}^{\infty}(k+1)^{n-1} \nonumber \\  
\times ~ \mathrm{exp}\bigg\{\frac {-(k+1)E } {T}\bigg\} {\mathrm{exp}\bigg\{ 
\frac{(k+1)\mu}{T}\bigg\}},
\label{eq:susc_mes}
 \end{eqnarray}
 and for baryons,
\begin{eqnarray}
\chi_{x,baryon}^{(n)}=\frac{X^n}{VT^3} \int{d^{3}p}\sum_{k=0}^{\infty}{(-1)^k}
(k+1)^{n-1} \nonumber \\ 
\times ~ \mathrm{exp}\bigg\{\frac{-(k+1)E} {T}\bigg\} 
{\mathrm{exp}\bigg\{\frac{(k+1)\mu}{T}\bigg\}},
\label{eq:susc_bar}
\end{eqnarray}
where $X$ represents either $B_i$, $Q_i$ or $S_i$ of the $i$-th particle 
depending on whether the susceptibility $\chi_{x}$ represents for net-baryon, 
net-electric charge or net-strangeness. The total generalized susceptibilities 
will be the sum of susceptibility of mesons and baryons as 
$\chi_x^{(n)} = \sum \chi_{x,mesons}^{(n)} + \sum \chi_{x,baryons}^{(n)}$.

In the HRG model, at the chemical freeze-out time, all the particles 
(primordial as well as resonances) are in equilibrium. The collision energy 
dependence of freeze-out parameters ($\mu_B$ and $T_f$) are parametrized as 
given in~\cite{Cleymans:2005xv}. The energy dependence of $\mu_B$ is given as, 
$\mu_B(\sqsn) = 1.308$ / (1 $+$ 0.273  $\sqsn$) and the $\mu_B$ dependence of 
freeze-out temperature is given as $T_f(\mu_B) =$ 0.166 $-$ 0.139 $\mu_B^2 -$ 
0.053 $\mu_B^4$. Further, the ratio of baryon to strangeness chemical potential 
on the freeze-out curve is parametrized as $\frac{\mu_S}{\mu_B} \simeq$ 0.164 + 
0.018\sqsn. Similarly, the energy dependence of $\mu_Q$ and $\mu_S$ are 
parametrized as given in~\cite{Karsch:2010ck}. In the HRG model, after the 
production of all the particles, the resonances are allowed to decay to their 
corresponding daughter particles, hence contributing to the final abundance of 
the stable meson and baryon numbers. These decay daughters from the resonance 
can influence the fluctuation of the final hadrons. The ensemble averaged stable 
particle yield will have contribution from both primordial production and the 
resonance decay~\cite{Jeon:1999gr, Begun:2006jf},
\begin{equation}
     \langle N_i\rangle = \langle N_i^*\rangle + \sum_R \langle N_R \rangle 
\langle n_i\rangle_R
\label{eq:ave}
\end{equation}
where $\langle N_i^*\rangle$ and $\langle N_R \rangle$ correspond to the 
average primordial yield of particle species $i$ and of the resonances $R$, 
respectively. The summation runs over all the resonances which decay to the 
final particle $i$ and $\langle n_i \rangle_R = \sum_r b_r^R n_{i,r}^R$ is the 
average number of particle type $i$ produced from the resonance $R$. Further, 
$b_r^R$ is the branching ratio of the $r$-th decay channel of the resonance $R$ 
and $n_{i,r}^R$ is the number of particle $i$ produced in that decay branch. 
The generalized susceptibility for stable particle $i$ of $n$-th order can be 
written as 
\begin{equation}
     \chi_i^{(n)} = \chi_i^{*(n)} + \sum_R \chi_R^{(n)}\langle n_i\rangle ^n_R
     \label{eq:chi_ave}
\end{equation}
The second term in Eq.~\ref{eq:chi_ave} includes contributions from 
fluctuations of primordial resonances and average number of produced particle of 
type $i$. We follow Eqs. (17) - (20) 
from Ref.~\cite{Nahrgang:2014fza} to calculate the average contributions from 
the resonances, where we consider fluctuation in the resonance production only 
and the number of decay daughters are assumed to be fixed. In this work, we 
refer this as average decay contributions. For completeness we rewrite the same 
cumulant equations here as given in Ref.~\cite{Nahrgang:2014fza};
\begin{equation}
      \langle (\Delta N_i)^2\rangle = \langle (\Delta N_i^*)^2\rangle + \sum_R 
\langle (\Delta N_R)^2 \rangle \langle n_i\rangle^2_R
\label{eq:c2}
\end{equation}
\begin{equation}
      \langle (\Delta N_i)^3\rangle = \langle (\Delta N_i^*)^3\rangle + \sum_R 
\langle (\Delta N_R)^3 \rangle \langle n_i\rangle^3_R
\label{eq:c3}
\end{equation}
\begin{equation}
      \langle (\Delta N_i)^4\rangle = \langle (\Delta N_i^*)^4\rangle + \sum_R 
\langle (\Delta N_R)^4 \rangle \langle n_i\rangle^4_R
\label{eq:c4}
\end{equation}
Resonance decay processes are probabilistic in nature, which itself 
causes the final particle number fluctuations. In above 
Eqs.~(\ref{eq:c2})--(\ref{eq:c4}), we use $\langle n_i\rangle$ which is same as 
the sum of the branching ratios of the different decay branches of the 
resonance. But the number of decay products follow a random distribution which 
gives fluctuation in the final number of particles. Detailed discussion of 
resonance decay is given in 
Refs.~\cite{Begun:2006jf,Fu:2013gga,Nahrgang:2014fza}. After considering the 
fluctuation in the produced daughters, the modified cumulants in 
Eqs.~(\ref{eq:c2})--(\ref{eq:c4}) of the stable particle from the resonance 
contributions (full decay) can be written as \cite{Nahrgang:2014fza};
\begin{eqnarray}
    \langle (\Delta N_i)^2\rangle &=& \langle (\Delta N_i^*)^2\rangle + \sum_R 
\langle (\Delta N_R)^2 \rangle \langle n_i\rangle^2_R \nonumber\\ 
&+&\sum_R \langle N_R \rangle\langle(\Delta n_i)^2\rangle_R
\label{eq:c2f}
\end{eqnarray}
\begin{eqnarray}
    \langle (\Delta N_i)^3\rangle &=& \langle (\Delta N_i^*)^3\rangle + \sum_R 
\langle (\Delta N_R)^3 \rangle \langle n_i\rangle^3_R \nonumber \\
 &+& 3\sum_R (\langle \Delta N_R)^2 \rangle\langle n_i \rangle_R \langle(\Delta 
n_i)^2\rangle_R \nonumber\\ &+& \sum_R \langle N_R \rangle \langle (\Delta 
n_i)^3\rangle_R
\label{eq:c3f}
\end{eqnarray}
\begin{eqnarray}
    \langle (\Delta N_i)^4\rangle &=& \langle (\Delta N_i^*)^4\rangle + \sum_R 
\langle (\Delta N_R)^4 \rangle \langle n_i\rangle^4_R \nonumber \\
&+& 6\sum_R \langle (\Delta N_R)^3 \rangle\langle n_i \rangle_R^2 
\langle(\Delta n_i)^2\rangle_R \nonumber\\ 
&+& \sum_R \langle (\Delta N_R)^2 \rangle  \bigg[ 3\langle 
(\Delta n_i)^2\rangle_R^2 \nonumber \\
&+& 4\langle n_i \rangle_R \langle (\Delta n_i)^3\rangle_R \bigg] \nonumber \\
&+&\sum_R \langle N_R \rangle \langle(\Delta n_i)^4\rangle_R
\label{eq:c4f}
\end{eqnarray}
Above Eqs.~(\ref{eq:c2f})--(\ref{eq:c4f}), which we refer as full decay, the 
fluctuation of the daughter particles are also considered along with the 
fluctuation of the resonances. If there is no correlation among the 
daughter particles, the fluctuation in the multiplicity calculated using 
Eqs.~(\ref{eq:c2f})--(\ref{eq:c4f}) will be very close to the average 
fluctuation contribution Eqs.~(\ref{eq:ave})--(\ref{eq:c4}) from resonance 
decay~\cite{Nahrgang:2014fza}. 
The higher order terms $\langle (\Delta n_i)^2\rangle_R$, $\langle 
(\Delta n_i)^3\rangle_R$, and $\langle (\Delta n_i)^4\rangle_R$ will be zero for 
the resonances having only one decay channel or the number of species $i$ is 
same in each decay branch. Hence, the higher order terms will have higher 
contribution for net-charge and net-kaon case with compared to net-proton. In 
this calculation, we have included mesons and baryons of mass up to 2.5 GeV as 
listed in the particle data book. We consider two different cases, one with 
weakly decaying particles regarded as stable and in another letting the weakly 
decaying particles decay into stable ones. In line with 
Ref.~\cite{Nahrgang:2014fza}, for the first case, we have considered 26 weakly 
decaying particles as stable. By including weak decays in addition to the 
strongly decaying particles in our analysis, we observe substantial change on 
the values of cumulant ratios for all the conserved number which we will 
discuss in the following section. 

%%%%%%%%%%%%%%%%%%%%%% Fig.1 %%%%%%%%%%%%%%%%%%%%%%%%%%%%%%%%%
\bef[ht!]
\bc
\includegraphics[width=0.5\textwidth]{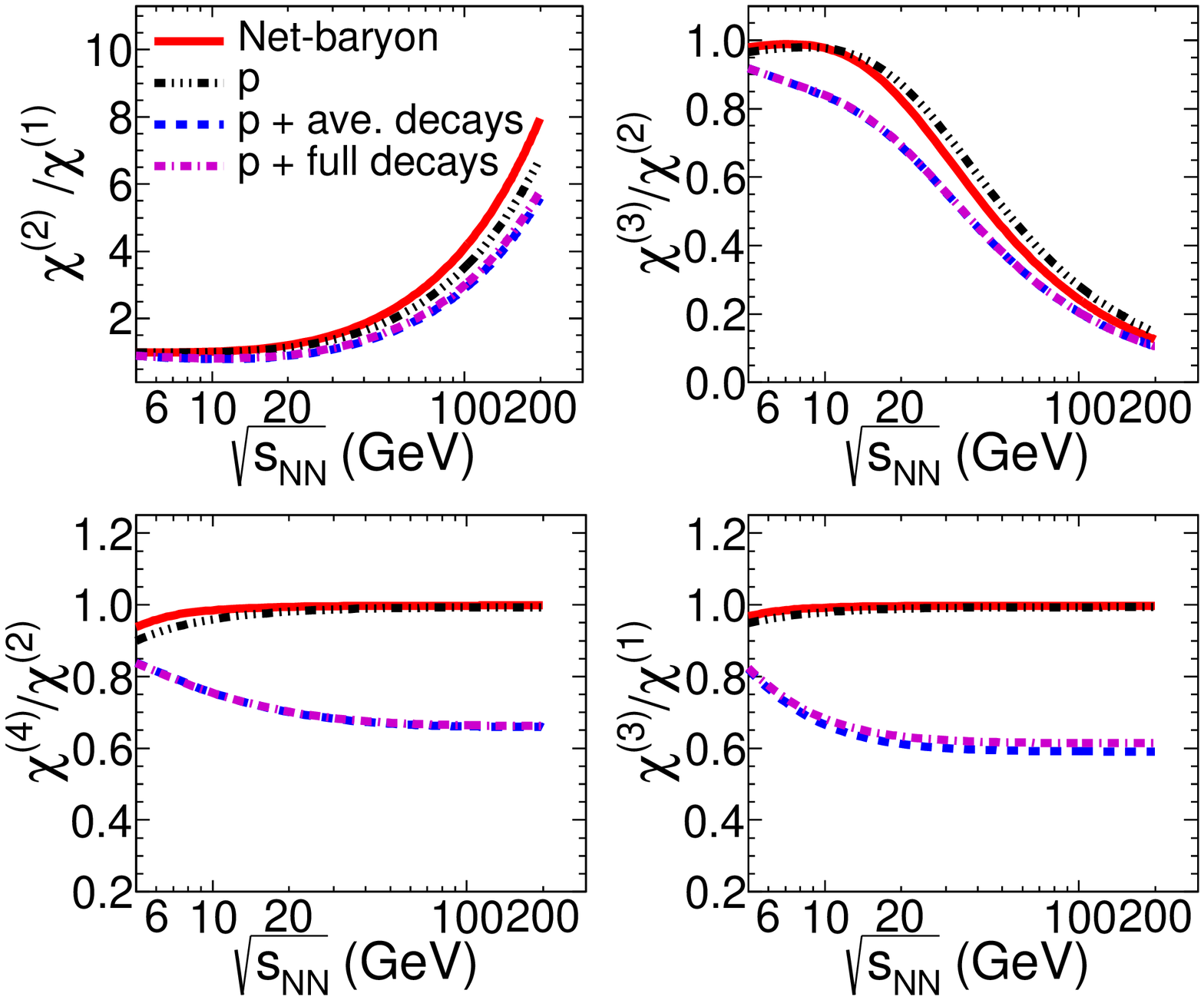}
\caption{Collision energy dependence of susceptibility ratios 
($\chi^{(2)}/\chi^{(1)}$, $\chi^{(3)}/\chi^{(2)}$, $\chi^{(4)}/\chi^{(2)}$ and 
$\chi^{(3)}/\chi^{(1)}$) calculated in full phase space for net-baryons without 
resonance decay, primordial protons and primordial protons with resonance decay 
including weak decay resonances. }
\label{fig:netb_decay_3}     
\ec
\eef
%%%%%%%%%%%%%%%%%%%%%%%%%%%%%%%%%%%%%%%%%%%%%%%%%%%%%%%%%%%%%%%%%
%%%%%%%%%%%%%%%%%%%%%% Fig.2 %%%%%%%%%%%%%%%%%%%%%%%%%%%%%%%%%
\bef[ht!]
\bc
\includegraphics[width=0.5\textwidth]{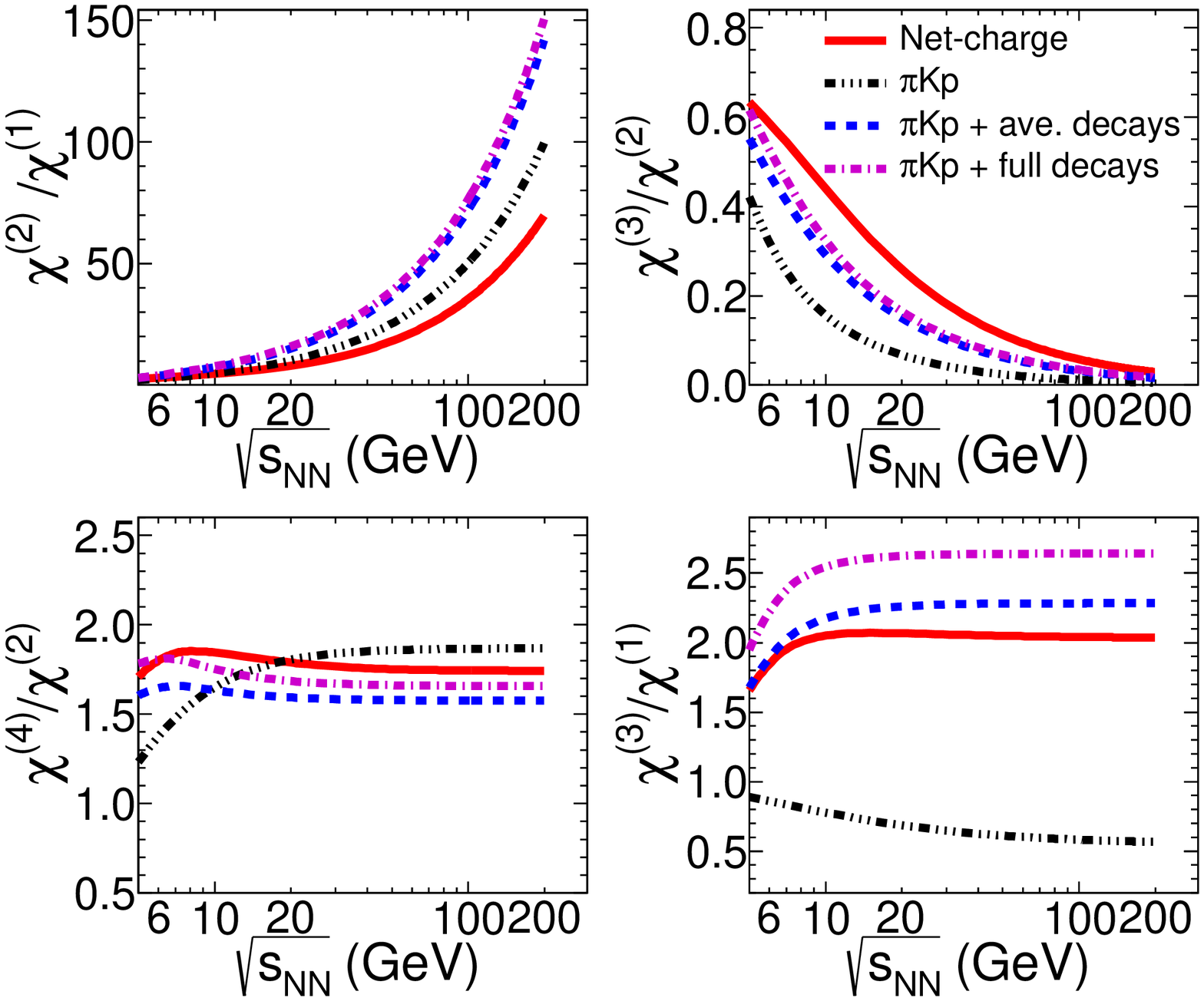}
\caption{Collision energy dependence of susceptibility ratios 
($\chi^{(2)}/\chi^{(1)}$, $\chi^{(3)}/\chi^{(2)}$, $\chi^{(4)}/\chi^{(2)}$, and 
$\chi^{(3)}/\chi^{(1)}$) calculated in full phase space for net-charge without 
resonance decay, primordial charged particles ($\pi, K, p$) and primordial 
charged hadrons with resonance decay including weak decay resonances.}
\label{fig:netq_decay_3}  
\ec
\eef
%%%%%%%%%%%%%%%%%%%%%%%%%%%%%%%%%%%%%%%%%%%%%%%%%%%%%%%%%%%%%%%%%
%%%%%%%%%%%%%%%%%%%%%% Fig.3 %%%%%%%%%%%%%%%%%%%%%%%%%%%%%%%%%
\bef[ht!]
\bc
\includegraphics[width=0.5\textwidth]{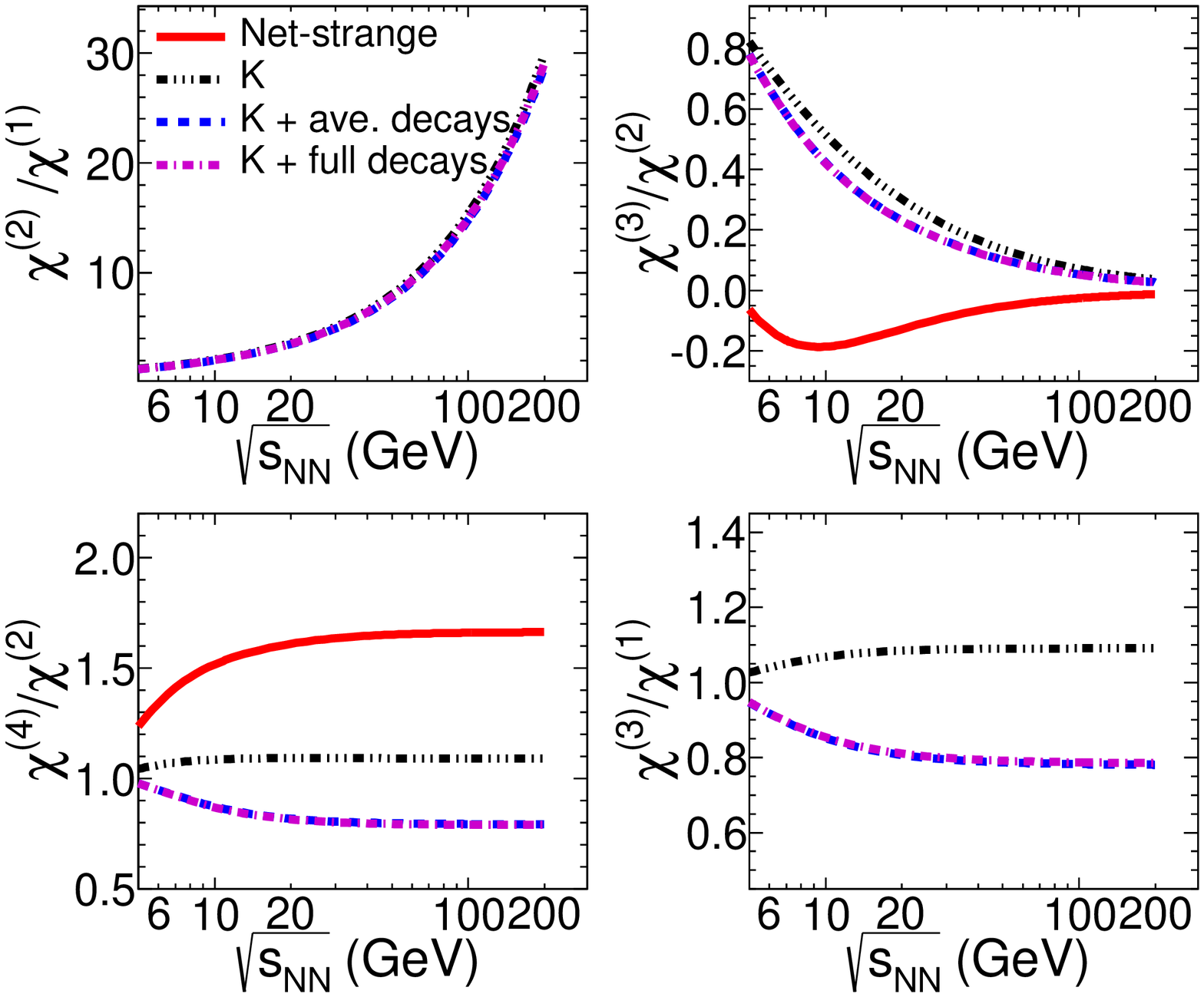}
\caption{Collision energy dependence of susceptibility ratios 
($\chi^{(2)}/\chi^{(1)}$, $\chi^{(3)}/\chi^{(2)}$, $\chi^{(4)}/\chi^{(2)}$ and 
$\chi^{(3)}/\chi^{(1)}$) calculated in full phase space for net-strangeness 
without resonance decay, primordial kaons and primordial kaons with resonance 
decay including weak decay resonances.}
\label{fig:nets_decay_3} 
\ec
\eef
%%%%%%%%%%%%%%%%%%%%%%%%%%%%%%%%%%%%%%%%%%%%%%%%%%%%%%%%%%%%%%%%%
\section{Results and discussion}
\label{sec:results}
Figures~(\ref{fig:netb_decay_3})--(\ref{fig:nets_decay_3}) show the variation 
of the susceptibility ratios for net-baryon, net-charge and net-strangeness as 
a function of the collision energies by considering average decay 
(Eqs.~(\ref{eq:ave})--(\ref{eq:c4})) and full decay contributions 
(Eqs.~(\ref{eq:c2f})--(\ref{eq:c4f})) of resonances as discussed in previous 
section. Figure~\ref{fig:netb_decay_3} shows ratios considering all 
the baryons including the resonances without decay, only primordial protons, 
primordial protons with average decay contributions from baryonic 
resonances using Eqs.~(\ref{eq:ave})--(\ref{eq:c4}) and primordial protons 
with contributions from fluctuation of resonances and their daughter particles 
(full decay) using Eqs.~(\ref{eq:c2f})--(\ref{eq:c4f}). 
Experimentally net-baryon fluctuations are accessible through 
net-proton fluctuations. There is significant effect of decay contributions 
observed with compared to no decay of resonances. However, the difference 
between average decay and full decay is negligible, which is in agreement with 
the findings of Ref.~\cite{Nahrgang:2014fza}. 

The net-charge fluctuations are accessible through measurement of fluctuation 
of stable charged particles ($\pi$, $K$ and $p$). Figure~\ref{fig:netq_decay_3} 
shows susceptibility ratios for net-charge which includes all the resonances 
without decay, only primordial stable charged particles ($\pi$, $K$ and $p$), 
primordial particles with average decay contributions from the 
resonances using Eqs.~(\ref{eq:ave})--(\ref{eq:c4}) and primordial stable 
particles with full decay contributions using 
Eqs.~(\ref{eq:c2f})--(\ref{eq:c4f}). There is substantial change 
in the susceptibility ratios by including the resonance decay contributions, 
particularly for higher \sqsn of $\chi^{(2)}/\chi^{(1)}$ and 
$\chi^{(3)}/\chi^{(2)}$ ratios for lower collision energies. The resonance 
decay effect for net-charge is larger as compared to net-baryon, because in case 
of baryons all the baryonic resonances decay into only one baryon in each decay 
branch, which is not the case for net-charge. Most of the higher mass resonances 
decay into more than one charged particle. Also the higher mass resonances again 
decay into resonance which after few decay iterations decay into final stable 
hadrons. Further, the neutral resonances also contribute to the net-charge 
fluctuation. For example, $\rho^0$ meson which decays into $\pi^+\pi^-$ about 
100\%, if both the daughter particles will be accepted in the detector then the 
contribution to the mean of the net-charge from $\rho^0$ will be zero. In an 
ideal HRG model in the grand canonical ensemble, thermally produced and 
non-interacting particles and anti-particles are uncorrelated, hence the 
susceptibility of the net-conserved quantity is : $\chi^{(n)}_{net} = 
\chi^{(n)}_{+} + (-1)^n\chi^{(n)}_{-}$. 
%%%%%%%%%%%%%%%%%%%%%% Fig.4 %%%%%%%%%%%%%%%%%%%%%%%%%%%%%%%%%
\bef[ht]
\bc
\includegraphics[width=0.5\textwidth]{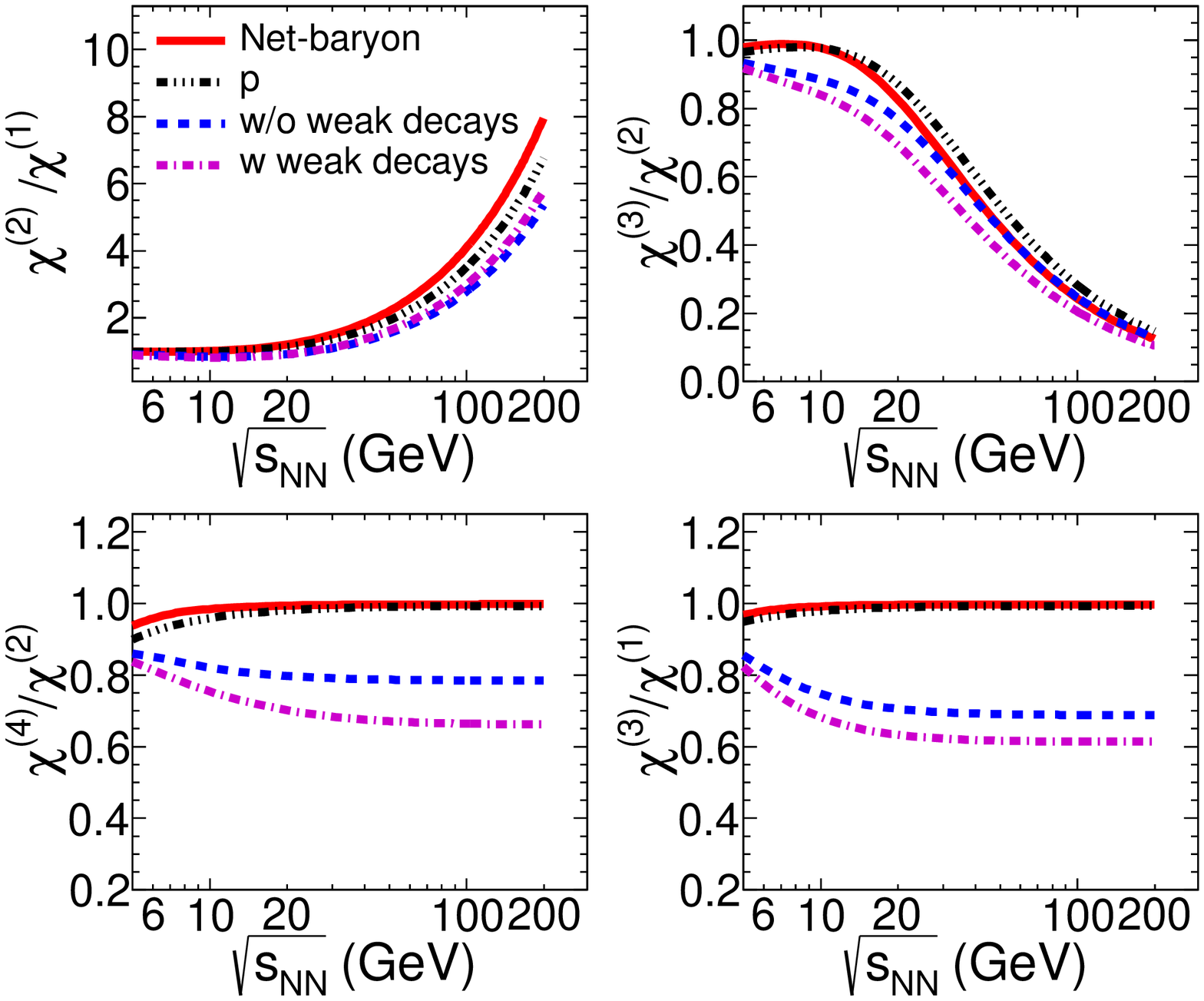}
\caption{Variation of susceptibility ratios ($\chi^{(2)}/\chi^{(1)}$, 
$\chi^{(3)}/\chi^{(2)}$, $\chi^{(4)}/\chi^{(2)}$ and $\chi^{(3)}/\chi^{(1)}$) 
as a function of \sqsn calculated in full phase space for net-baryon without 
resonance decay, primordial proton, primordial proton with and without 
including weak decay in addition to the strongly decaying resonances.}
\label{fig:netb_decay_sta} 
\ec
\eef
%%%%%%%%%%%%%%%%%%%%%%%%%%%%%%%%%%%%%%%%%%%%%%%%%%%%%%%%%%%%%%%%%
%%%%%%%%%%%%%%%%%%%%%% Fig.5 %%%%%%%%%%%%%%%%%%%%%%%%%%%%%%%%%
\bef[ht!]
\bc
\includegraphics[width=0.5\textwidth]{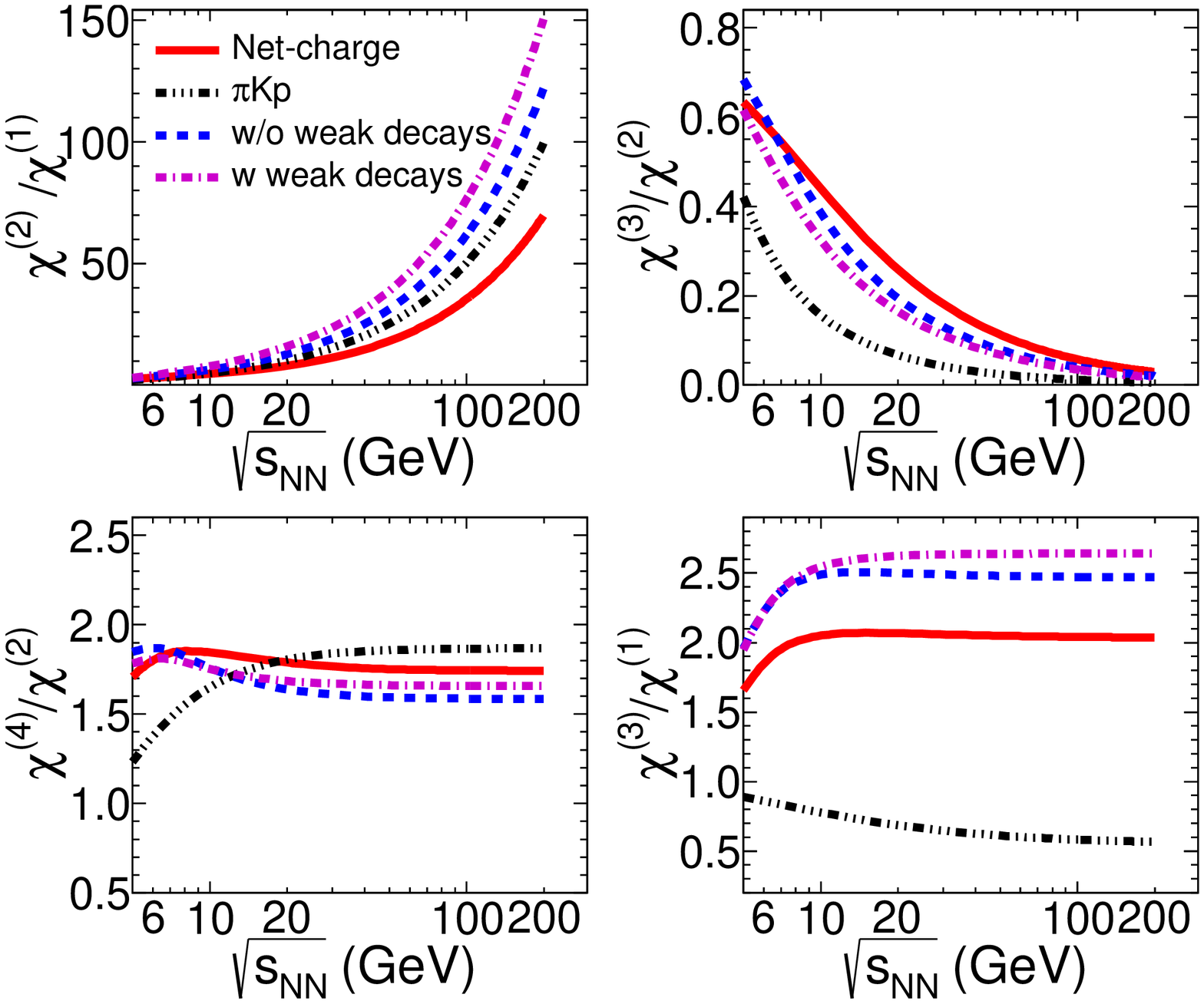}
\caption{Variation of susceptibility ratios ($\chi^{(2)}/\chi^{(1)}$, 
$\chi^{(3)}/\chi^{(2)}$, $\chi^{(4)}/\chi^{(2)}$ and $\chi^{(3)}/\chi^{(1)}$) 
as a function of \sqsn calculated in full phase space for net-charge without 
resonance decay, primordial charged particles ($\pi, K, p$), primordial charged 
particles with and without including weak decay in addition to the strongly 
decaying resonances.}
\label{fig:netq_decay_sta}
\ec
\eef
%%%%%%%%%%%%%%%%%%%%%%%%%%%%%%%%%%%%%%%%%%%%%%%%%%%%%%%%%%%%%%%%%
%%%%%%%%%%%%%%%%%%%%%% Fig.6 %%%%%%%%%%%%%%%%%%%%%%%%%%%%%%%%%
\bef[ht!]
\bc
\includegraphics[width=0.5\textwidth]{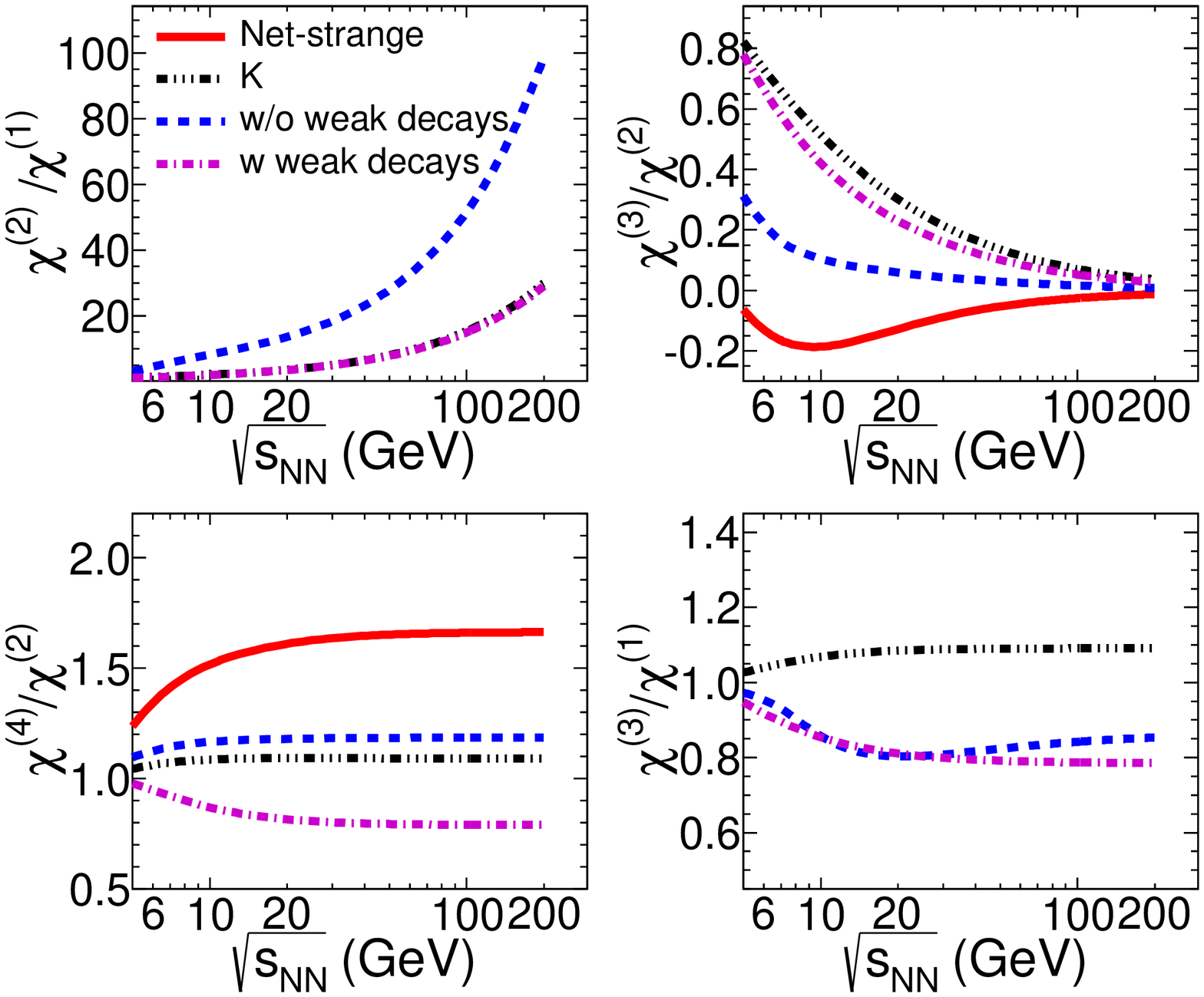}
\caption{Variation of susceptibility ratios ($\chi^{(2)}/\chi^{(1)}$, 
$\chi^{(3)}/\chi^{(2)}$, $\chi^{(4)}/\chi^{(2)}$ and $\chi^{(3)}/\chi^{(1)}$) 
as a function of \sqsn calculated in full phase space for net-strange without 
resonance decay, primordial kaons, primordial kaons with and without including 
weak decay in addition to the strongly decaying resonances.}
\label{fig:nets_decay_sta}
\ec
\eef
%%%%%%%%%%%%%%%%%%%%%%%%%%%%%%%%%%%%%%%%%%%%%%%%%%%%%%%%%%%%%%%%%
The second and fourth order susceptibilities of $\rho^0$ will contribute to the 
net-charge susceptibility as the susceptibility of particle ($\pi^+$) and 
anti-particle ($\pi^-$) will get added. But the contribution of $\rho^0$ to the 
first and third order susceptibilities of the net-charge will be zero. This may 
be one reason why there is more effect of resonance decay in 
$\chi^{(2)}/\chi^{(1)}$ and $\chi^{(3)}/\chi^{(2)}$ ratios. 

Experimentally, the net-strangeness fluctuations are accessible through 
measuring the net-kaon 
fluctuations. Figure~\ref{fig:nets_decay_3} shows the susceptibility ratios of 
net-strange particles without resonance decay, considering only primordial 
kaons, primordial kaons with average decay contributions from the 
strange resonances using Eqs.~(\ref{eq:ave})--(\ref{eq:c4}) and primordial 
kaons with full decay contributions using Eqs.~(\ref{eq:c2f})--(\ref{eq:c4f}). 
For net-strangeness also there is significant resonance 
contributions observed in $\chi^{(3)}/\chi^{(2)}$ and $\chi^{(4)}/\chi^{(2)}$ 
ratios from the resonance decay with compared to no decay of resonances. We 
would like to mention here that, the mean of the net-strangeness multiplicity is 
zero due to the imposed strangeness neutrality and iso-spin asymmetry in the 
initial state of Au $+$ Au collisions~\cite{Karsch:2010ck}. Therefore, 
$\chi^{(2)}/\chi^{(1)}$ and $\chi^{(3)}/\chi^{(1)}$ diverges in case of 
net-strange multiplicity distributions in HRG model, which are not shown in the 
Fig.~\ref{fig:nets_decay_3}.

Figures~(\ref{fig:netb_decay_sta})--(\ref{fig:nets_decay_sta}) show the energy 
dependence of susceptibility ratios for net-baryon, net-charge and 
net-strangeness. Figure~\ref{fig:netb_decay_sta} shows the susceptibility ratios 
for net-baryon without resonance decay, primordial net-protons, net-protons from 
primordial and resonance contributions without weak decays and including weak 
decay particles in addition to the resonance contributions from strong decay. 
The resonance decay contributions are calculated using full decay contributions 
using Eqs.~(\ref{eq:c2f})--(\ref{eq:c4f}). We consider $K^0$, $\bar 
K^0$, $\eta^0$, $\Lambda^0$, $\Sigma^{\pm}$, $\Sigma^0$, $\Xi^-$, $\Xi^0$,  
$\Omega^-$ and their anti-particles as weak decay particles. In 
Ref.~\cite{Nahrgang:2014fza}, these particles were considered as stable 
particles and only strongly decaying particles were considered in the resonance 
decay. In the present work we have explicitly considered their weak decays in 
addition to the strongly decaying particles. There is very small effect in 
$\chi^{(2)}/\chi^{(1)}$ ratio for all the energies, but the 
$\chi^{(3)}/\chi^{(2)}$, $\chi^{(4)}/\chi^{(2)}$ and 
$\chi^{(3)}/\chi^{(1)}$ ratios further decrease at all \sqsn as compared to the 
values of excluding weak decays. Similarly, Fig.~\ref{fig:netq_decay_sta}, shows 
the susceptibility ratios for net-charge without decay of resonances, 
primordial $\pi$, $K$, $p$, and contributions from strongly decaying resonances 
along with and without weak decays. There is visible difference between the 
results with and without inclusion of weak decays. 
Figure~\ref{fig:nets_decay_sta}, shows the susceptibility ratios for 
net-strangeness without decay of resonances, only 
primordial kaons, with and without inclusion of weak decays in addition to 
contribution from strongly decaying resonances. There is significant difference 
of the susceptibility ratios for with and without inclusion of weak decay 
contributions except $\chi^{(3)}/\chi^{(1)}$ ratio. For all the 
cases, net-baryon, net-charge and net-strangeness, it is important to consider 
weak decays contributions in addition to the strongly decaying particles while 
comparing model calculations with the experimental observables. Without decay 
of resonances and primordial contributions in 
Figs.~(\ref{fig:netb_decay_sta})--(\ref{fig:nets_decay_sta}) are same as shown 
in Figs.~(\ref{fig:netb_decay_3})--(\ref{fig:nets_decay_3}).

%%%%%%%%%%%%%%%%%%%%%% Fig.7 %%%%%%%%%%%%%%%%%%%%%%%%%%%%%%%%%
\bef[t]
\bc
\includegraphics[width=0.5\textwidth]{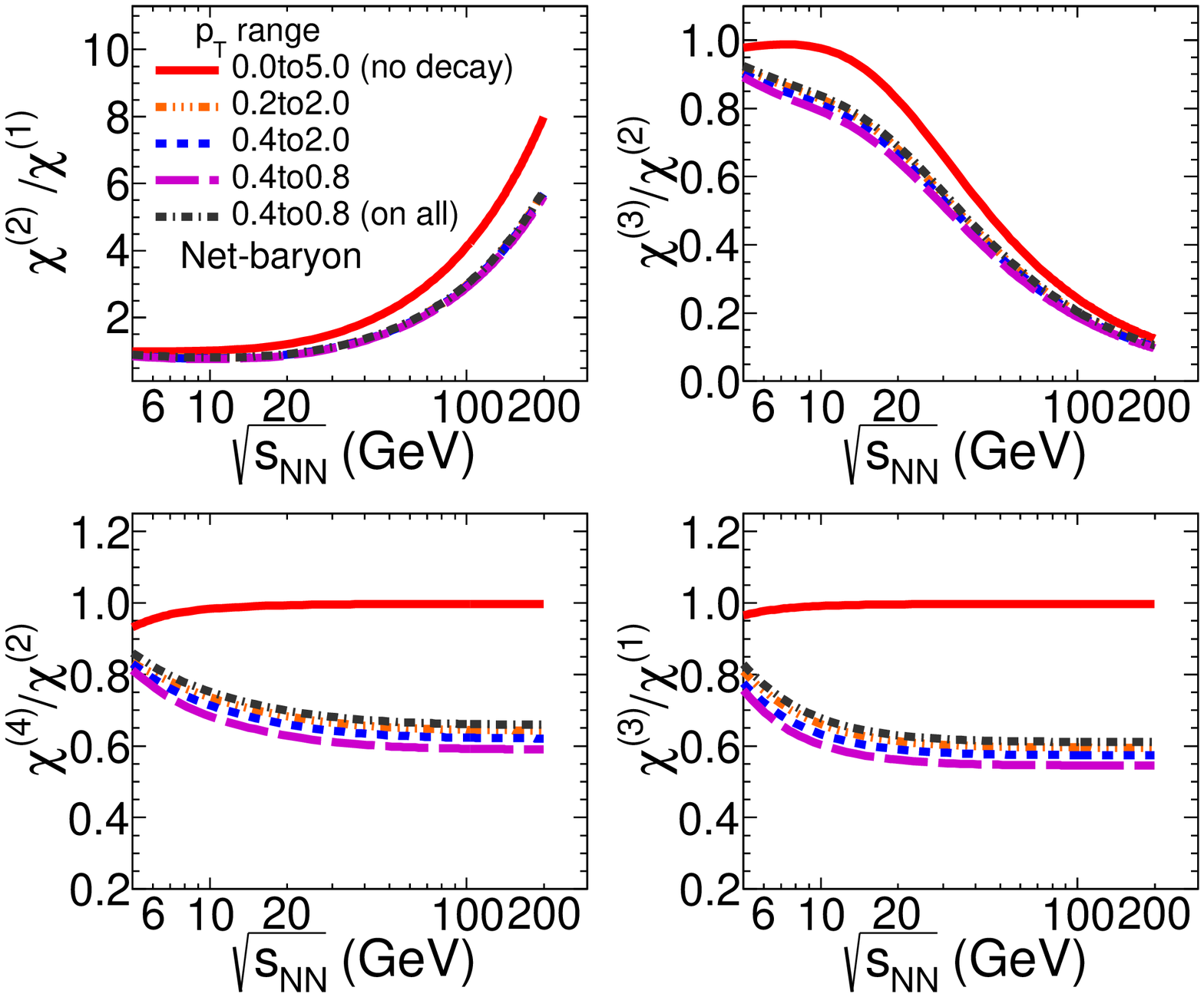}
\caption{Collision energy dependence of susceptibility ratios 
($\chi^{(2)}/\chi^{(1)}$, $\chi^{(3)}/\chi^{(2)}$, $\chi^{(4)}/\chi^{(2)}$ and 
$\chi^{(3)}/\chi^{(1)}$) for net-protons for different $p_T$ acceptances 
within $|\eta|<$ 0.5. The results are for primordial protons with resonance 
decay including weak decay resonances.}
\label{fig:netb_decay_pt}
 
\ec
\eef
%%%%%%%%%%%%%%%%%%%%%%%%%%%%%%%%%%%%%%%%%%%%%%%%%%%%%%%%%%%%%%%%%
%%%%%%%%%%%%%%%%%%%%%% Fig.8 %%%%%%%%%%%%%%%%%%%%%%%%%%%%%%%%%
\bef[ht!]
\bc
\includegraphics[width=0.5\textwidth]{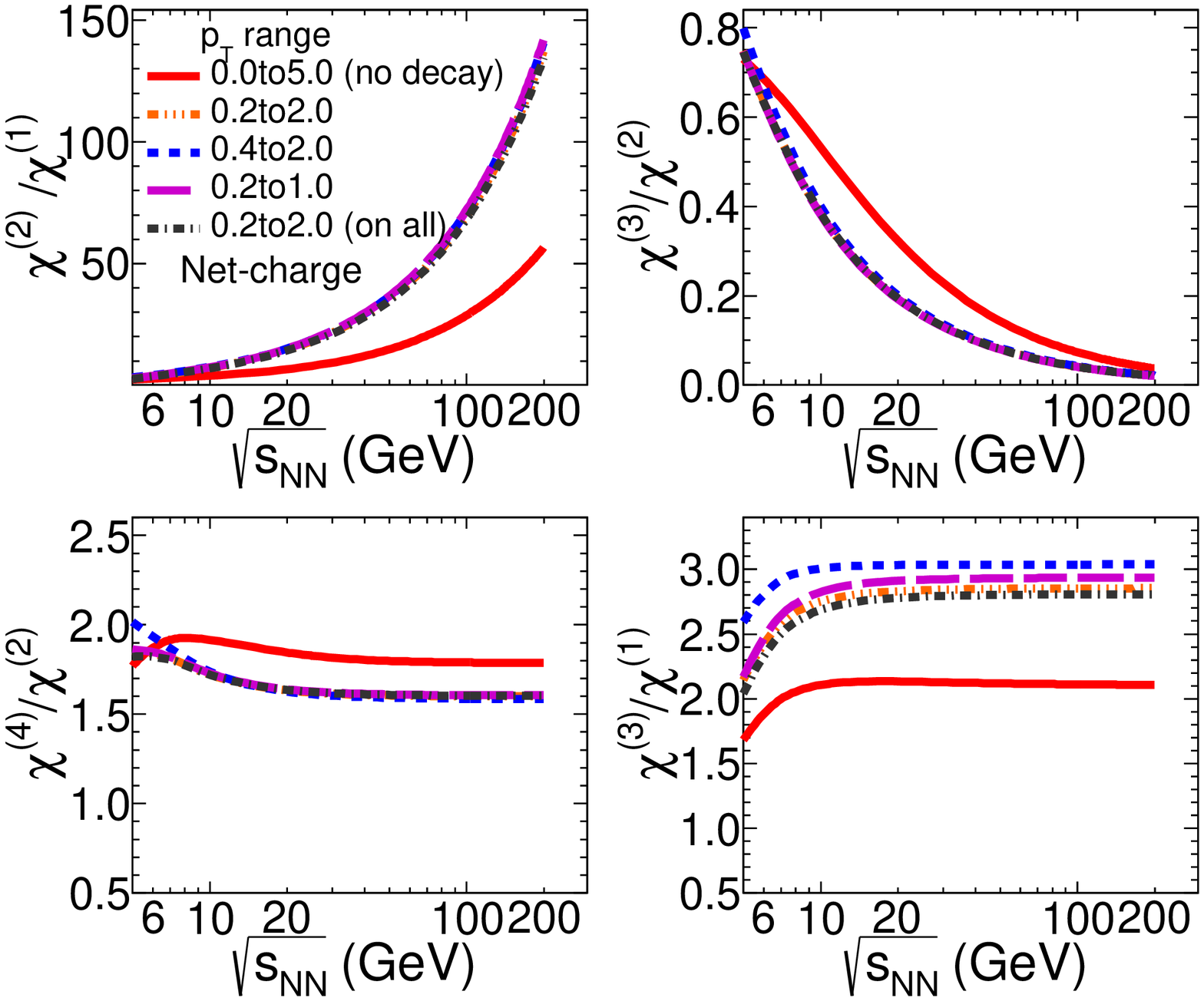}
\caption{Collision energy dependence of susceptibility ratios 
($\chi^{(2)}/\chi^{(1)}$, $\chi^{(3)}/\chi^{(2)}$, $\chi^{(4)}/\chi^{(2)}$ and 
$\chi^{(3)}/\chi^{(1)}$) for net-charge for different $p_T$ acceptances with 
$|\eta|<$ 0.5. The results are for primordial charged particles ($\pi, K, p$) 
with resonance decay including weak decay resonances.}
\label{fig:netq_decay_pt}
\ec
\eef
%%%%%%%%%%%%%%%%%%%%%%%%%%%%%%%%%%%%%%%%%%%%%%%%%%%%%%%%%%%%%%%%%

%%%%%%%%%%%%%%%%%%%%%% Fig.9 %%%%%%%%%%%%%%%%%%%%%%%%%%%%%%%%%
\bef[ht!]
\bc
\includegraphics[width=0.5\textwidth]{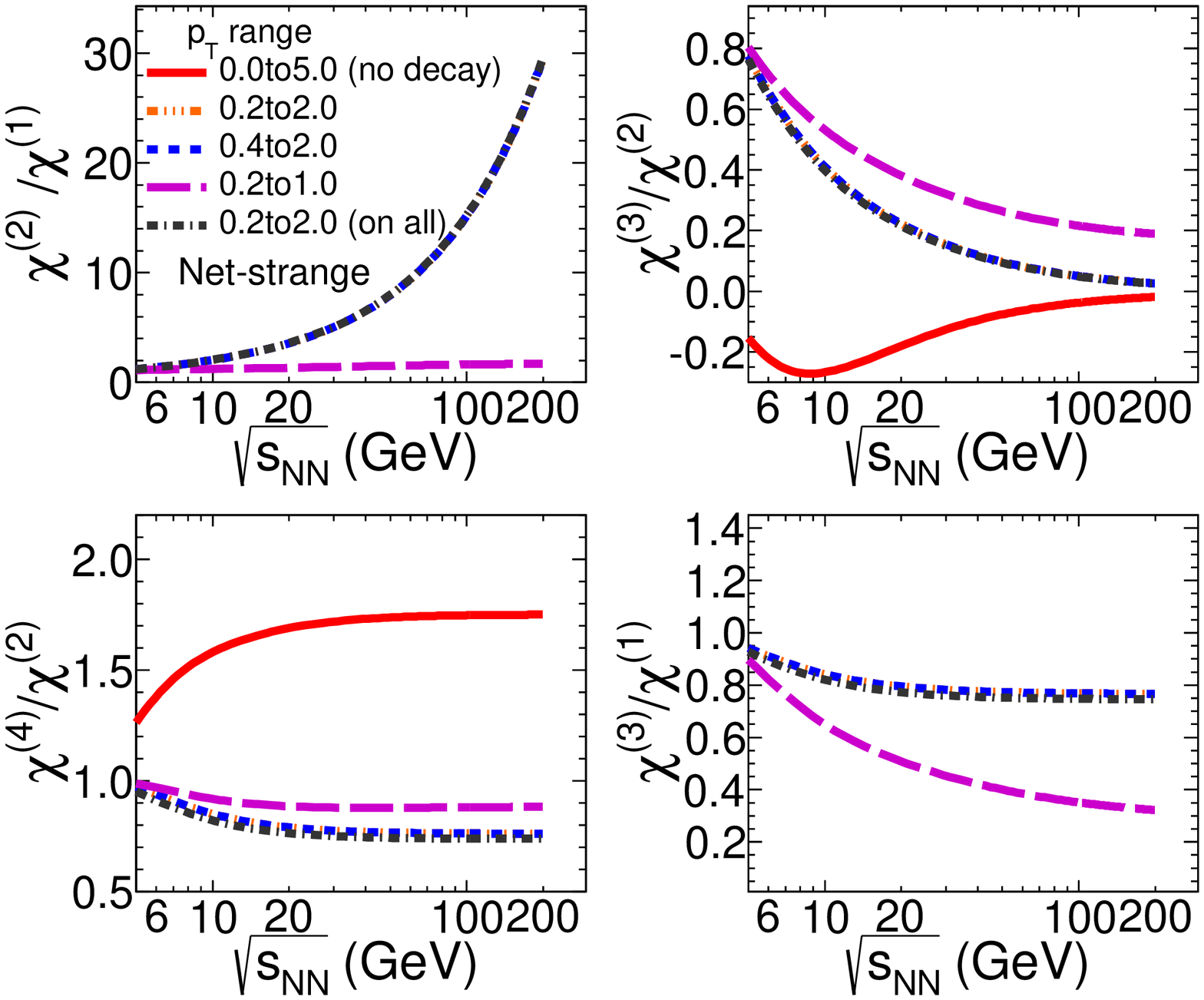}
\caption{Collision energy dependence of susceptibility ratios 
($\chi^{(2)}/\chi^{(1)}$, $\chi^{(3)}/\chi^{(2)}$, $\chi^{(4)}/\chi^{(2)}$ and 
$\chi^{(3)}/\chi^{(1)}$) for net-kaons for different $p_T$ acceptances within 
$|\eta<$ 0.5. The results are for primordial kaons with resonance decay 
including weak decay resonances.}
\label{fig:nets_decay_pt}
\ec
\eef
%%%%%%%%%%%%%%%%%%%%%%%%%%%%%%%%%%%%%%%%%%%%%%%%%%%%%%%%%%%%%%%%%

Figures~(\ref{fig:netb_decay_pt})--(\ref{fig:nets_decay_pt}) show the variation 
of susceptibility ratios as a function of \sqsn for various $p_T$ acceptances 
for net-baryon, net-charge and net-strangeness. The ratios without resonance 
decay are also shown for comparison. The $p_T$ acceptance cuts have been applied 
to the stable particles only and the resonances are taken in full $p_T$ range. 
We have also shown another case where $p_T$ cut is applied to all the particles 
(stable as well as resonances). Although there is significant difference between 
with and without resonance decay, the $p_T$ acceptance has very minimal effect 
on the susceptibility ratios after inclusion of resonance decay for all 
the conserved quantities. In Ref.~\cite{Garg:2013ata}, which was studied 
without taking resonance decay into account, a clear $p_T$ dependence was 
observed for net-charge and net-strangeness cases at all collision energies. In 
reality the acceptance cuts should be applied separately on the daughters of 
resonances. The resonance may be produced in full acceptance which can be 
outside the experimental acceptance, still the decay daughters have a chance to 
be accepted within the experimental acceptance because of their decay 
kinematics. It has been mentioned in Ref.~\cite{Nahrgang:2014fza} that, due to 
the elastic scatterings in thermally equilibrated hadronic phase, the kinematic 
cuts affect the same manner for the primordial particle and anti-particle from 
the resonance decay. However, this may not be true when the detector will have 
asymmetric azimuthal acceptance.

\section{Comparison to the experimental measurements}
\label{sec:compare}
%%%%%%%%%%%%%%%%%%%%%% Fig.10 %%%%%%%%%%%%%%%%%%%%%%%%%%%%%%%%%
\bef[ht!]
\bc
\includegraphics[width=0.47\textwidth]{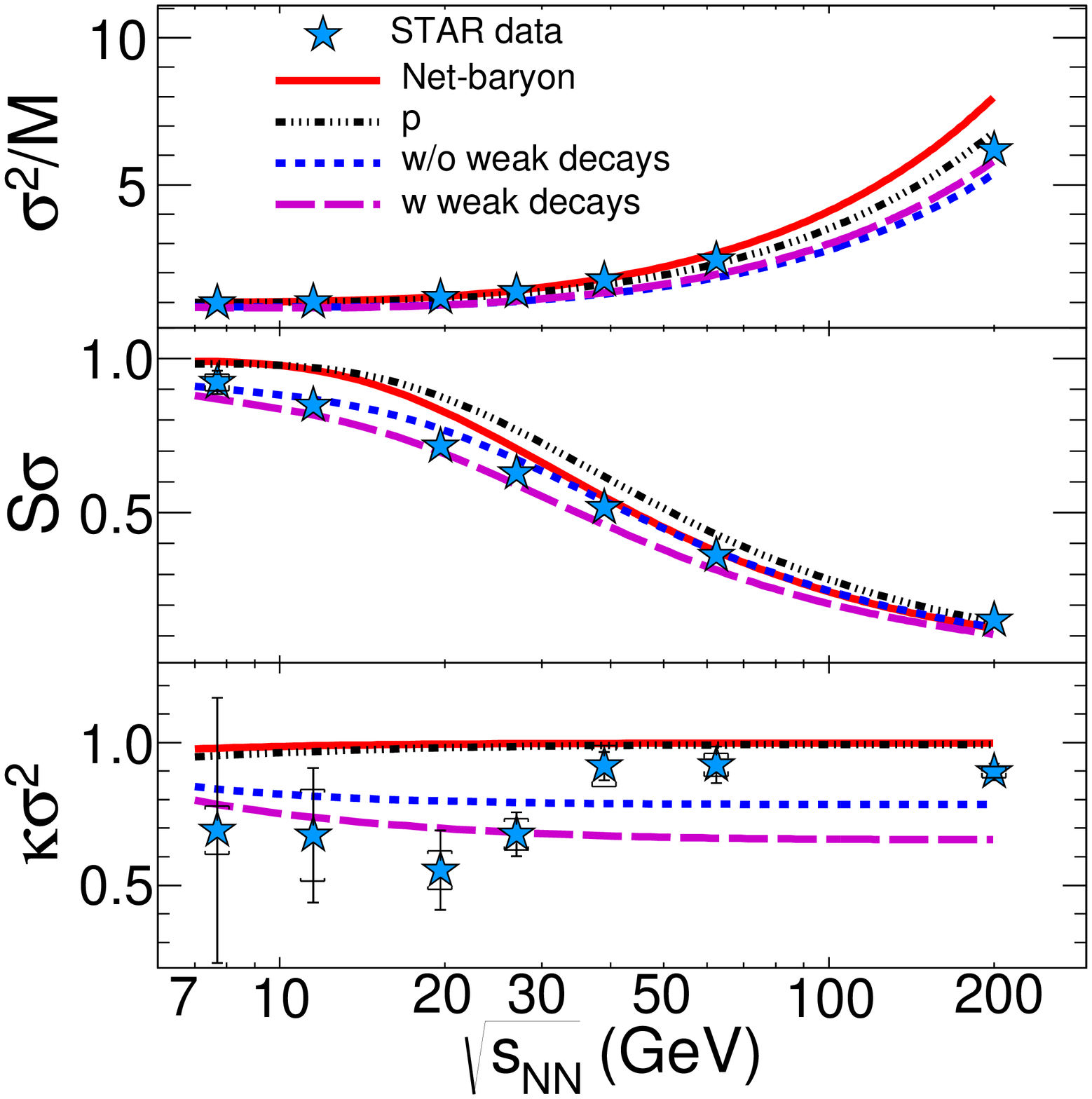}
\caption{The collision energy dependence of $\sigma^2/M$, $S\sigma$, and 
$\kappa\sigma^2$ of the net-baryon calculated using HRG model including 
resonance decay contributions. The model calculations are compared with the 
experimental net-proton cumulant ratios for most central (0--5)\% Au $+$ Au 
collisions by STAR experiment.}
\label{fig:compare_starp}
\ec
\eef
%%%%%%%%%%%%%%%%%%%%%%%%%%%%%%%%%%%%%%%%%%%%%%%%%%%%%%%%%%%%%%%%%
Experimentally measured moments ($M$, $\sigma$, $S$, $\kappa$) of the net 
distributions are related to the susceptibilities as: $\sigma^2/M \sim 
\chi^{(2)}/\chi^{(1)}$, $S\sigma \sim \chi^{(3)}/\chi^{(2)}$, $\kappa\sigma^2 
\sim \chi^{(4)}/\chi^{(2)}$. Figure~\ref{fig:compare_starp} shows the energy 
dependence of $\sigma^2/M$, $S\sigma$ and $\kappa\sigma^2$ of net-proton 
distribution for central (0--5)\% Au $+$ Au collisions measured by STAR 
experiment~\cite{Adamczyk:2013dal}. The experimental data is studied within 
mid-rapidity ($|y| <$ 0.5) and $p_T$ range 0.4 to 0.8 GeV/$c$. The data is 
compared with the HRG calculations with no decay of resonances, only 
primordial protons, resonance decay with and without inclusion of weak decay 
contributions in addition to the contribution from strongly decaying resonances 
within the same experimental acceptance. We would like to mention 
here that, we have applied same $p_T$ acceptance cuts to the primordial as well 
as resonances. The HRG calculations without resonance decay fail to explain 
$\sigma^2/M$ at higher collision energies and $S\sigma$ values at lower $\sqsn$. 
The $\sigma^2/M$ values are well described by considering only primordial 
protons but over estimates the $S\sigma$ and $\kappa\sigma^2$ values. The 
$\sigma^2/M$ calculated in HRG with resonance decay along with inclusion of weak 
decay contributions are closer to the experimental values. The $S\sigma$ values 
lies between resonance decays with and without inclusion of weak decays. The 
$\kappa\sigma^2$ values at lower \sqsn are better described by inclusion of weak 
decays. For higher collision energy (above 30 GeV), the HRG model under predict 
the experimental values. This may be because of the regeneration of the 
resonances at higher collision energies as observed in 
Refs.~\cite{Nahrgang:2014fza,Kitazawa:2012at}. Hence it is important to 
consider the weak decays in addition to the strongly decaying resonances in the 
HRG model while comparing with experimental data.

%%%%%%%%%%%%%%%%%%%%%% Fig.11 %%%%%%%%%%%%%%%%%%%%%%%%%%%%%%%%%
\bef[ht]
\bc
\includegraphics[width=0.45\textwidth]{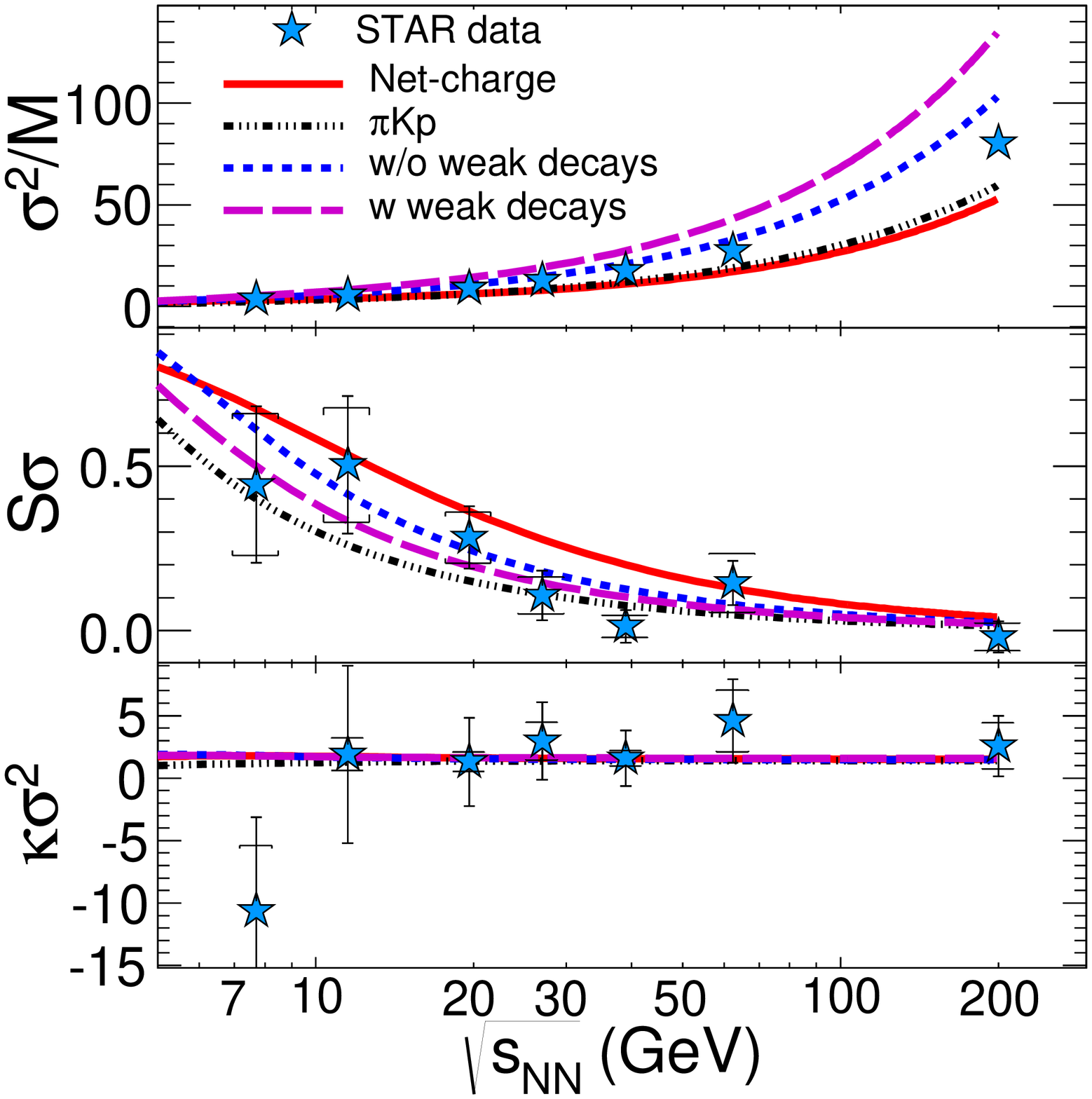}
\caption{The collision energy dependence of $\sigma^2/M$, $S\sigma$, and 
$\kappa\sigma^2$ of the net-charge calculated using HRG model including 
resonance decay contributions. The model calculations are compared with the 
experimental net-charge cumulant ratios for most central (0--5)\% Au $+$ Au 
collisions by STAR experiment.}
\label{fig:compare_starq}
\ec
\eef
%%%%%%%%%%%%%%%%%%%%%%%%%%%%%%%%%%%%%%%%%%%%%%%%%%%%%%%%%%%%%%%%%

Figure~\ref{fig:compare_starq} shows the energy dependence of $\sigma^2/M$, 
$S\sigma$ and $\kappa\sigma^2$ of net-charge distributions for most central 
(0--5)\% Au $+$ Au collisions within $|\eta| <$ 0.5 and $p_T$ range within 0.2 
to 2.0 GeV/$c$ measured by STAR experiment~\cite{Adamczyk:2014fia}. The 
experimental net-charge results are compared with the HRG calculations 
considering without resonance decay, only charged stable particles ($\pi, K, 
p$), resonance decay with and without inclusion of weak decays along with 
strongly decaying resonances. The HRG results for $\sigma^2/M$ without resonance 
decay and considering only primordial particles shows lower values with compared 
to the experimental data. Where as the HRG calculations with resonance decay 
over estimates the experimental data. Inclusion of weak decays further worsen 
the agreement with the experimental data at higher \sqsn. The experimental 
$S\sigma$ and $\kappa\sigma^2$ values are explained by all cases of HRG because 
of the large uncertainties in the measured experimental data.

%%%%%%%%%%%%%%%%%%%%%% Fig.12 %%%%%%%%%%%%%%%%%%%%%%%%%%%%%%%%%
\bef[ht!]
\bc
\includegraphics[width=0.45\textwidth]{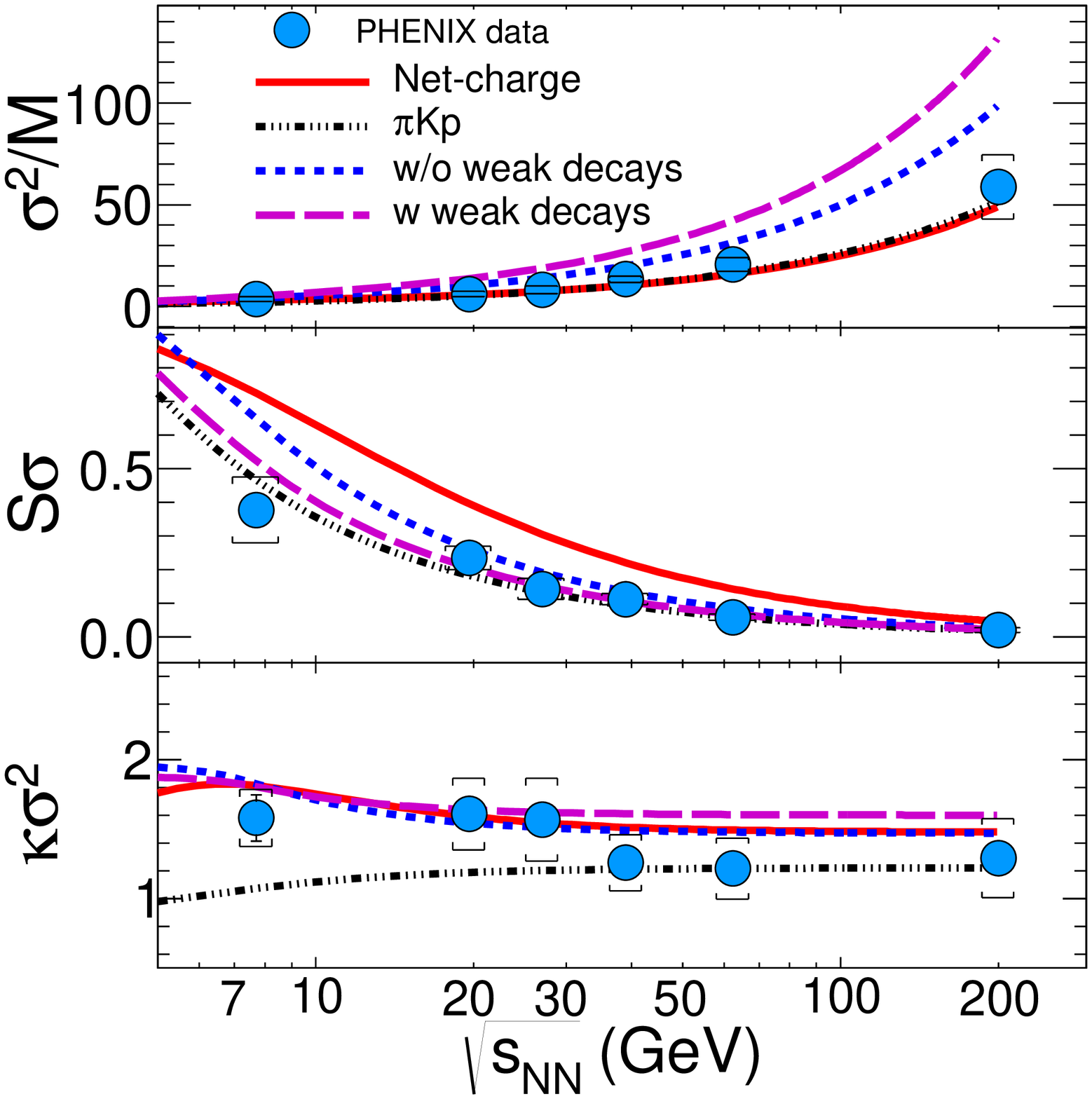}
\caption{The collision energy dependence of $\sigma^2/M$, $S\sigma$, and 
$\kappa\sigma^2$ of the net-charge calculated using HRG model including 
resonance decay contributions. The model calculations are compared with the 
experimental net-charge cumulant ratios for most central (0--5)\% Au $+$ Au 
collisions by PHENIX experiment.}
\label{fig:compare_phenixq}
\ec
\eef
%%%%%%%%%%%%%%%%%%%%%%%%%%%%%%%%%%%%%%%%%%%%%%%%%%%%%%%%%%%%%%%%%

Figure~\ref{fig:compare_phenixq} shows the collision energy dependence of 
$\sigma^2/M$, $S\sigma$ and $\kappa\sigma^2$ of net-charge distribution for 
most central bin (0--5)\% in Au $+$ Au collisions measured by PHENIX 
experiment~\cite{Adare:2015aqk}. The experimental measurements are within 
$|\eta| < $ 0.35 and $p_T$ between 0.3 -- 2.0 GeV/$c$. The experimental data 
are compared with the HRG calculations considering no decay of resonances, only 
primordial charged hadrons ($\pi, K, p$), with and without inclusion of weak 
decays in the resonance decay. The kinematic acceptance cuts applied to the 
HRG  calculations are same as in experimental data. As observed in STAR 
net-charge results, the HRG calculations without resonance decay  are more 
close to the experimental $\sigma^2/M$ values. Inclusion of resonance decay 
over estimate the  experimental data. Inclusion of weak decay in the resonance 
decay further deviates from the experimental values. 
The disagreement in $\sigma^2/M$ between experimental data and the HRG may be 
because of the acceptance cuts. As mentioned before, we have applied the 
kinematic acceptance cuts on the resonance not on their daughter particles. 
Although, if we consider the resonances in full phase space for a $4\pi$ 
detector then the neutral resonances will not contribute to the net-charge 
fluctuations. For example the case $\rho^0$, $K^{*0}$ or $\Lambda^0$, if we 
measure the daughter particles in full phase space, they will also not 
contribute to the net-charge fluctuation. This may be one of the reason, why 
$\sigma^2/M$ are not explained by HRG model with resonance decay. The HRG 
calculations without resonance decay fails to explain the experimental $S\sigma$ 
at all \sqsn. The $S\sigma$ are well explained by resonance decay with weak 
decay contributions. In this case also, results from only primordial charged 
particles are more close to the experimental data. The $\kappa\sigma^2$ values 
are well explained by HRG with resonance decay within the experimental 
uncertainties at all the studied energies. The $\kappa\sigma^2$ values without 
resonance decay also shows similar results as with resonance decay. However, 
considering only primordial charged particles explain the experimental data 
very well at higher \sqsn but fails to explain at lower collision energies.

\section{Summary}
\label{sec:summary}
In conclusion we have studied the effect of resonance decay on conserved 
number fluctuations using a hadron resonance gas model. There is a significant 
effect of resonance decay as compared to no decay of resonances for all the 
conserved number fluctuations. The effect of considering primordial particles 
with average decay contributions of the resonances are studied. There is small 
effect whether we consider average decay or full decay of resonances.  The 
inclusion of weak resonance decays in addition to the strongly decaying 
particles show visible difference compared to without inclusion of weak decays. 
The effect of different $p_T$ acceptance cuts on the resonance decay are very 
minimal for all the conserved numbers. The experimental data for net-proton and 
net-charge are compared with the HRG calculations, which are estimated within 
the similar kinematic acceptance as in the experiment. The cumulant ratios of 
net-proton distributions are better explained by considering the resonance 
decay contributions and the agreement further improves by inclusion of weak 
resonance decays. The HRG calculations for net-charge with resonance decay over 
estimates the experimental $\sigma^2/M$ values. However, the $S\sigma$ and 
$\kappa\sigma^2$ of net-charge experimental values are well explained by HRG 
calculations with resonance decay, which further improves by including the weak 
decays. Hence, it is important to consider resonance decays and with weak decay 
resonance contribution before comparing the model calculations with experimental 
data.

% Non-BibTeX users please use

\end{document}